%
% Multiple-trace interactions and non-local string theories
%
%Last edited by OA, 7/12

\input harvmac.tex
\noblackbox

\newcount\yearltd\yearltd=\year\advance\yearltd by 0

\input epsf
\newcount\figno
\figno=0
\def\fig#1#2#3{
\par\begingroup\parindent=0pt\leftskip=1cm\rightskip=1cm\parindent=0pt
\baselineskip=11pt
\global\advance\figno by 1
\midinsert
\epsfxsize=#3
\centerline{\epsfbox{#2}}
\vskip 12pt
{\bf Figure \the\figno:} #1\par
\endinsert\endgroup\par
}
\def\figlabel#1{\xdef#1{\the\figno}}

%\draftmode

%
\def\np#1#2#3{Nucl. Phys. {\bf B#1} (#2) #3}
\def\plb#1#2#3{Phys. Lett. {\bf #1B} (#2) #3}

\def\physrev#1#2#3{Phys. Rev. {\bf D#1} (#2) #3}

\def\prep#1#2#3{Phys. Rep. {\bf #1} (#2) #3}

\def\mpl#1#2#3{Mod. Phys. Lett. {\bf A#1} (#2) #3}

\def\atmp#1#2#3{Adv. Theor. Math. Phys. {\bf #1} (#2) #3}
\lref\ls{R. G. Leigh and M. J. Strassler, ``Exactly marginal operators
and duality in four dimensional $\cn=1$ supersymmetric gauge theory,''
hep-th/9503121, \np{447}{1995}{95}.}%

\lref\juan{J. M. Maldacena, ``The large $N$ limit of superconformal field
theories and supergravity", hep-th/9711200, \atmp{2}{1998}{231}.}

\lref\gkp{S. S. Gubser, I. R. Klebanov and A. M. Polyakov,
``Gauge theory correlators
from non-critical string theory", hep-th/9802109,
\plb{428}{1998}{105}.}

\lref\witten{E. Witten, ``Anti-de-Sitter space and holography",
hep-th/9802150, \atmp{2}{1998}{253}.}

\lref\review{O. Aharony, S. S. Gubser, J. Maldacena, H. Ooguri and Y. Oz,
``Large $N$ field theories, string theory and gravity,'' hep-th/9905111,
\prep{323}{2000}{183}.}

\lref\tsezar{
A.~A.~Tseytlin and K.~Zarembo,
``Effective potential in non-supersymmetric $SU(N) \times SU(N)$
gauge theory  and interactions of type $0$ D3-branes,''
Phys.\ Lett.\ B {\bf 457}, 77 (1999),
hep-th/9902095.}

\lref\adsil{
A.~Adams and E.~Silverstein,
``Closed string tachyons, AdS/CFT, and large $N$ QCD,''
hep-th/0103220.}

\lref\klewit{I.~R.~Klebanov and E.~Witten,
``Superconformal field theory on threebranes at a Calabi-Yau  singularity,''
\np{536}{1998}{199}, hep-th/9807080.}

\lref\gubser{S.~S.~Gubser, ``Einstein manifolds and conformal field
theories,'' hep-th/9807164, \physrev{59}{1999}{025006}.}

\lref\klegub{
S.~S.~Gubser and I.~R.~Klebanov,
``Baryons and domain walls in an $\cn = 1$ superconformal gauge theory,''
hep-th/9808075, \physrev{58}{1998}{125025}.}

\lref\igor{I.~R.~Klebanov, ``Touching random surfaces and Liouville
gravity,'' \physrev{51}{1995}{1836}, hep-th/9407167.}

\lref\igoraki{I. R. Klebanov and A. Hashimoto,
``Nonperturbative solution of matrix models modified by trace
squared terms'', Nucl. Phys. B {\bf 434}, 264 (1995), hep-th/9409064.}

\lref\gk{
D. Kutasov and D. A. Sahakyan, ``Comments on the thermodynamics
of Little String Theory'', JHEP {\bf 0102}, 021 (2001), hep-th/0012258.
}
\lref\willy{W.~Fischler, S.~Paban and M.~Rozali,
``Collective coordinates for D-branes,''
Phys.\ Lett.\ B {\bf 381}, 62 (1996),
hep-th/9604014.}

\lref\morpless{
D.~R.~Morrison and M.~R.~Plesser,
``Non-spherical horizons. I,''
Adv.\ Theor.\ Math.\ Phys.\  {\bf 3}, 1 (1999),
hep-th/9810201.
}

\lref\ddsw{S.~R.~Das, A.~Dhar, A.~M.~Sengupta, and S.~R.~Wadia, ``New
critical behavior in $d=0$ large $N$ matrix models,'' \mpl{5}{1990}{1041}.}

\lref\sugtsu{F.~Sugino and O.~Tsuchiya, ``Critical behavior in $c=1$
matrix model with branching interactions,'' \mpl{9}{1994}{3149},
hep-th/9403089.}

\lref\gubkle{S.~S.~Gubser and I.~R.~Klebanov, ``A modified $c=1$
matrix model with new critical behavior,'' \plb{340}{1994}{35},
hep-th/9407014.}

\lref\coleman{S. Coleman, ``$1/N$,'' in the proceedings of the 1979
Erice School on Subnuclear Physics and in ``Aspects of Symmetry,''
Cambridge University Press.}

%\BalasubramanianSN
\lref\BalasubramanianSN{
V.~Balasubramanian, P.~Kraus and A.~E.~Lawrence,
``Bulk vs. boundary dynamics in anti-de Sitter spacetime,''
Phys.\ Rev.\ D {\bf 59}, 046003 (1999),
hep-th/9805171.
%%CITATION = HEP-TH 9805171;%%
}

%\GiddingsQU
\lref\GiddingsQU{
S.~B.~Giddings,
``The boundary S-matrix and the AdS to CFT dictionary,''
Phys.\ Rev.\ Lett.\  {\bf 83}, 2707 (1999)
hep-th/9903048.
%%CITATION = HEP-TH 9903048;%%
}

%\PolchinskiRY
\lref\PolchinskiRY{
J.~Polchinski,
``S-matrices from AdS spacetime,''
hep-th/9901076.
%%CITATION = HEP-TH 9901076;%%
}

%\DavidVP
\lref\DavidVP{
F.~David,
``A scenario for the $c > 1$ barrier in non-critical bosonic strings,''
Nucl.\ Phys.\ B {\bf 487}, 633 (1997)
[hep-th/9610037].
%%CITATION = HEP-TH 9610037;%%
}

%\AndreevAY
\lref\AndreevAY{
O.~Andreev,
``On touching random surfaces, two-dimensional
quantum gravity and  non-critical string theory,''
Phys.\ Rev.\ D {\bf 57}, 3725 (1998)
[hep-th/9710107].
%%CITATION = HEP-TH 9710107;%%
}

%\ArutyunovIM
\lref\ArutyunovIM{
G.~Arutyunov, S.~Frolov and A.~Petkou,
``Perturbative and instanton corrections to the OPE of CPOs in $\cn = 4$
SYM(4),''
Nucl.\ Phys.\ B {\bf 602}, 238 (2001)
[hep-th/0010137].
%%CITATION = HEP-TH 0010137;%%
}

%\lref\twoparticle{
%}

%
\def\cn{{\cal N}}
\def\IZ{\relax\ifmmode\hbox{Z\kern-.4em Z}\else{Z\kern-.4em Z}\fi}
\def\IR{\relax{\rm I\kern-.18em R}}
\def\co{{\cal O}}
\def\th{{\tilde h}}

\Title{\vbox{\baselineskip12pt\hbox{hep-th/0105309}
\hbox{NSF-ITP-01-46}
\hbox{SLAC-PUB-8843}
\hbox{WIS/10/01-MAY-DPP}
}}
{\vbox{
{\centerline{Multiple-Trace Operators and}}
\vskip .1in
{\centerline{Non-Local String Theories}}
}}

\centerline{
Ofer Aharony$^{a,b,}$\foot{E-mail : {\tt Ofer.Aharony@weizmann.ac.il.}
Incumbent of the Joseph and Celia Reskin
career development chair.},
Micha Berkooz$^{a,b,}$\foot{E-mail : {\tt Micha.Berkooz@weizmann.ac.il.}},
and Eva Silverstein$^{b,c,}$\foot{E-mail : {\tt evas@slac.stanford.edu.}}}
\bigskip
\bigskip
\centerline{$^a${\it Dept. of Particle Physics,
The Weizmann Institute of Science, Rehovot 76100, Israel}}
\centerline{$^b${\it Institute for Theoretical
Physics, University of California,
Santa Barbara, CA 93106 USA}}
\centerline{$^c${\it Department of Physics and SLAC, Stanford University,
Stanford, CA 94305/94309 USA}}
\bigskip
\medskip
\noindent

We propose that a novel deformation of string perturbation theory,
involving non-local interactions between strings, is required to
describe the gravity duals of field theories deformed by
multiple-trace operators. The new perturbative expansion involves a
new parameter, which is neither the string coupling nor the
coefficient of a vertex operator on the worldsheet.  We explore some
of the properties of this deformation, focusing on a special case
where the deformation in the field theory is exactly marginal.

\Date{May 2001}

\newsec{Introduction}

Perturbative string theory provides us with a rigid set of rules for
computing S-matrix elements in particular limits of M Theory.  The
building blocks of these rules are the genus expansion of Riemann
surfaces, and a conformal field theory on each of these
surfaces. Although this is a very rich structure, which is only
partially understood, one wonders whether this is the most general set
of rules, or just the tip of the iceberg.  There are, of course, many
backgrounds for which we have no systematic perturbative description;
this is the case for generic backgrounds in M theory. There are also
decoupling
limits of M theory which give string theories which are inherently
strongly coupled, such as ``little string theories''.  However, all
previously studied backgrounds which are amenable to a perturbative
description can be described by the usual set of rules.

In this paper we will discuss new backgrounds of critical string theory,
which have a good perturbative description but which require
an enlarged set of rules. To construct these backgrounds we will use a
specific kinematical setting, that of the anti-de Sitter (AdS)/conformal
field theory (CFT) correspondence, which has already taught us many
surprising facts about string theory. This will allow us to give
a non-perturbative definition (and, in particular, a strong
argument for existence)
of these backgrounds. The usual set of rules involves
a fixed local conformal theory on the worldsheet and a perturbative
expansion in powers of the string coupling. In the enlarged rules we
will have an additional parameter, which does not correspond to the
string coupling or to a local
vertex operator on the worldsheet. One can think of the new parameter
as the weight of
new forms of degenerate worldsheets, which roughly
correspond to zero size worm-holes in the 2D gravity of the
worldsheet; this is analogous to the way in which the string coupling
is the weight of a handle on the worldsheet.
Alternatively, we can try
to sum the perturbation theory in the new parameter,
and then we remain with the usual genus
expansion but with a non-local worldsheet action, including interactions
between disconnected components of the worldsheet.

The kinematical context that we will be discussing is
a deformation of string theory on an
$AdS_5$ space, which is holographically dual to a four dimensional conformal
field theory \refs{\juan,\gkp,\witten,\review}. Deformations of
the conformal field theory by single-trace operators have been
extensively studied in the last few years, and we wish to generalize
the discussion to deformations of the conformal field theory by
multiple-trace operators. From the field theory point of view such
a deformation is not significantly different (at finite $N$)
from a single-trace deformation, but we
will see that from the point of view of string theory they seem quite
different. Our
main example will be based on type IIB string theory on
$AdS_5\times T^{1,1}$, which was discussed in
\refs{\klewit,\gubser,\klegub,\morpless}.
The corresponding ${\cal N}=1$ CFT is relatively well understood, and one can
show that it contains an exactly marginal deformation which is a
superpotential term involving a product of two gauge-invariant
chiral operators (whose
dimensions add up to 3). Phrased in the language of an asymptotically
free UV theory which flows to this CFT, the operator we deform by
can be written as
a product of two traces.  Double-trace perturbations can also be
generated radiatively as in the examples of \refs{\tsezar,\adsil}, where there
is again a family of possible coefficients determined in these
cases by a choice of renormalization-group trajectory in the
field theory.

In string theory on AdS space this presents
the following puzzle : on the one hand one expects to be able to
deform the string theory background
to include the double-trace deformation, but on
the other hand, there is no obvious parameter in conventional string
theory corresponding to such a deformation.  The usual parameters of
string theory involve turning on vertex operators on the worldsheet,
which in AdS is the same as changing the VEV of a field in spacetime;
but this corresponds to the deformation of the field theory by a
single-trace operator (or a simple generator of the chiral ring).
Note that, as frequently done in discussions of string theory in RR
backgrounds, we are assuming that the relevant conformal field theories
behave in a standard way, as described above,
though they are not well-understood.

Therefore, it is clear that these examples
with double-trace operators (or, more generally, multiple-trace operators)
appearing in the
Lagrangian lead to a prediction for a novel form of
perturbative string theory on the gravity side. In this paper we will
explain some aspects of the new perturbative expansion, and some of
its surprising features in the bulk of the target space. We will see
that the resulting string theories are non-local both on the
worldsheet and in space-time, so we dub them ``non-local string
theories'' (NLSTs).

We should emphasize that the role of {\it states} involving
multiple-trace operator
excitations around the ordinary unperturbed AdS/CFT background
is well known: they describe multiparticle states on the
gravity side (see, e.g., \review), and the anomalous dimensions of
the operators creating these states have been computed in some
examples. Our goal
here is to articulate properties of the gravity side of the
correspondence when we
{\it perturb the Lagrangian} by multiple-trace operators.

Double-trace perturbations of matrix models for non-critical strings
were studied in e.g. \refs{\ddsw,\sugtsu,\gubkle,\igor,\igoraki,
\DavidVP,\AndreevAY},
where it was observed that the presence of such terms seems to
lead to contact interactions on the string
worldsheet. In the matrix model context it was conjectured that these
terms may be interpreted
as changing the branch of the Liouville dressing; the relation between
this case and the critical string case we discuss here is not
clear.  The question  of what a multi-trace deformation
would mean on the gravity side of AdS/CFT dual pairs was raised
previously in \ArutyunovIM.

The organization of the paper is the following. We begin in section 2
by discussing general properties of field theories with
double-trace (and multiple-trace)
deformations in the 't Hooft large $N$ expansion.
In section 3 we
discuss the details of some specific examples where multiple-trace
perturbations arise on the field theory side of an AdS/CFT dual pair.
Using the field theory results, we
proceed in section 4 to give a worldsheet description of the new
``genus-wormhole'' expansion in a perturbative expansion around the
undeformed system, and discuss some of the features of the deformation
from the point of view of the gravitational dual theory, such as its
non-locality.
We also discuss briefly how
one might approach a more generic description of this type of background.
In section 5 we discuss an alternative approach to the problem, in
which we view the deformation as changing aspects of
boundary conditions on AdS.

Although we will exhibit the non-local deformation only for certain
AdS/CFT dual pairs, it raises the interesting question of whether this
sort of string perturbation theory might exist in (and define) more
general backgrounds. Such more general backgrounds could lead to new
low-energy effective actions. It is clear that the deformation does
not depend on the presence of RR background fields, since it exists
also in $AdS_3$ cases with NS-NS backgrounds, which we hope to study
in future work. However, it is not clear what are the requirements on
the asymptotic geometry for this type of deformation to make sense.
It should also be interesting to study the effects of the non-local
behavior we identify here on the calculation of
UV-sensitive quantities on the gravity side such as the vacuum
energy, and on non-local operators such as Wilson lines on the
field theory side.  Because
of the novelty of this sort of system, our analysis here is rather
preliminary, but we hope that our observations will help to stimulate
further development of these theories, and, perhaps, the construction
of a more general picture of what types of perturbative string
theories exist (and appear in some corners of the M theory moduli
space).

\newsec{The large $N$ limit of double-trace operators in field theory}

$SU(N)$ gauge theories in which all fields are in the adjoint representation
(or in bifundamental representations) usually have a Lagrangian
involving only terms which can be written as a single trace, like the
standard gauge kinetic term $\tr(F_{\mu \nu}^2)$. If one normalizes
the fields such that the Lagrangian is proportional to ${1\over
g_{YM}^2}$, there exists a 't Hooft large $N$ limit in which $\lambda
\equiv g_{YM}^2 N$ is kept constant. In this limit the perturbation
theory becomes a double expansion in $1/N^2$ and in
$\lambda$ (see, e.g. \coleman).
Diagrams which have genus $g$ (in the standard double line
notation for the Feynman diagrams) contribute with a factor of
$N^{2-2g}$ times some power of $\lambda$. In particular, in a standard
normalization for the single-trace operators $\co_i$ in the theory, in
which $\co_i$ is $N$ times a trace of a product of the fields (up to
some function of $\lambda$), the correlation functions of $\co_i$ have an
expansion of the form
\eqn\correl{\vev{\co_1 \co_2 \cdots \co_j} = \sum_{g=0}^{\infty}
N^{2-2g} f_g(\lambda).}
The first term in the sum, which corresponds to the planar diagrams,
dominates in the large $N$ limit. In \correl\ we only wrote down the
contribution from connected diagrams, of course there are also
disconnected diagrams whose generating function is the exponential of
the generating function for the connected diagrams.

\fig{Examples of vertices (in double line notation) corresponding to
(a) single-trace and (b) double-trace operators, where the indices
run over $1,\cdots,N$.}
{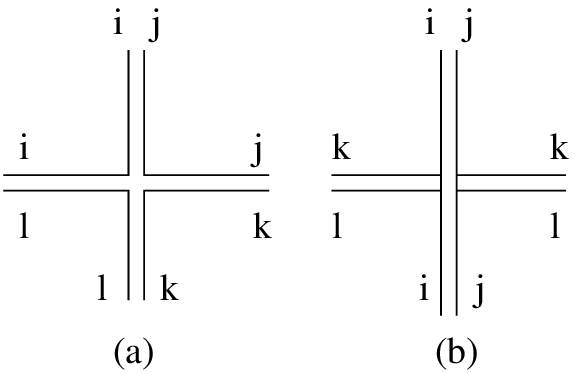}{10 truecm}
\figlabel\vertices

The building blocks of the diagrammatic expansion described above are
the single-trace vertices, which can be written in the plane as
described in figure \vertices (a). On the other hand, if we have a vertex
corresponding to a double-trace operator (such as $\tr(F_{\mu
\nu}^2)^2$), or an insertion of such an operator, then this cannot be
drawn in the plane, as in figure \vertices (b). Thus, naively
it seems that the
contribution of double-trace operators will always be negligible in
the large $N$ limit, since they should only appear in non-planar
diagrams.

\fig{A double-trace vertex connecting two components of a Feynman
diagram that was originally disconnected.}{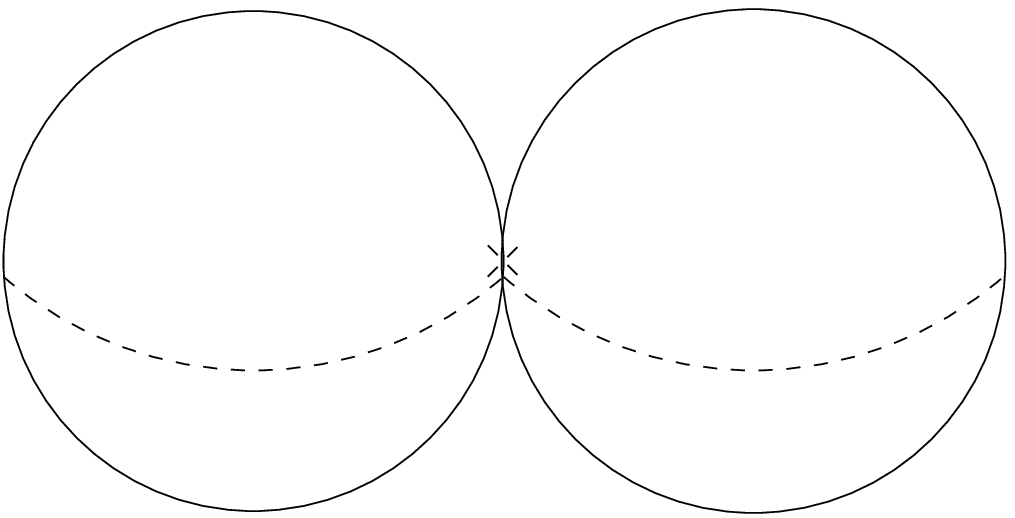}{8 truecm}
\figlabel\twospheres

However, there is also the possibility that the double-trace
operator can connect two parts of the Feynman diagram which were
originally disconnected, as in figure \twospheres. If instead of the
double-trace operator we had two single-trace operators (one in each
connected
component), whose coupling constant is taken to be proportional to $N$
in the 't Hooft large $N$ limit, then this
diagram would scale as $(N^2)^2$ (a product of two standard planar
contributions). Therefore, when we have the double-trace vertex
instead, such a diagram will scale as $N^2$ times the double-trace
coupling constant.

\fig{Various diagrams with standard vertices and double-trace
vertices. If $\th$ denotes the double-trace coupling, then
the diagram in the $i$'th row ($i=0,1,\cdots$) and the
$j$'th column ($j=0,1,\cdots$) will
scale as ${\th}^{i} N^{2-2j}$ in the large $N$ limit we are considering.}
{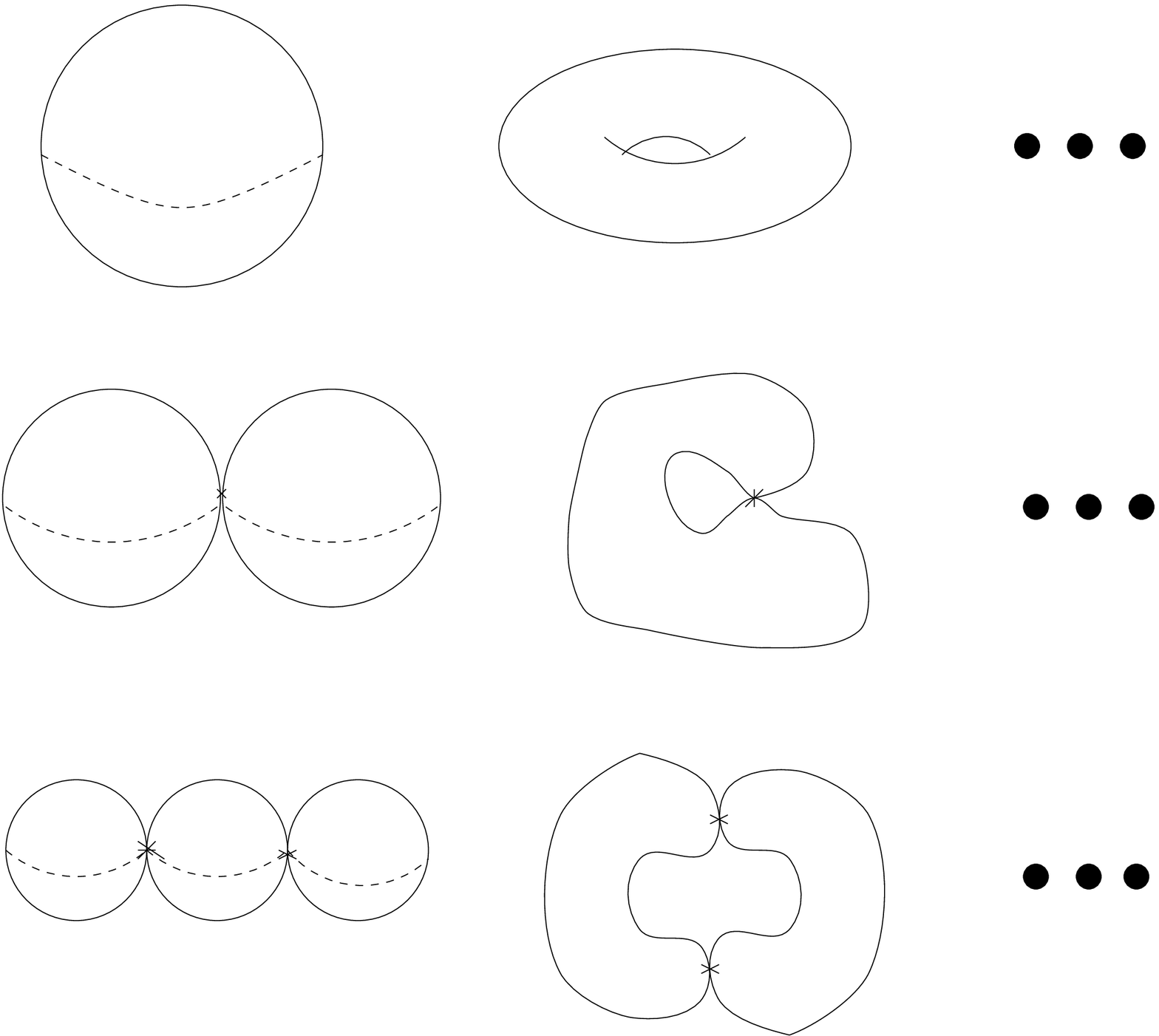}{12 truecm}
\figlabel\diagrams

If we scale the double-trace coupling constant such that it grows with
$N$ in the large $N$ limit, then this type of diagram, which becomes
disconnected when one removes the double-trace operators, will
dominate in the large $N$ limit, since its contribution would grow
faster than $N^2$. Such a large $N$ limit would not correspond to a
string theory, but perhaps to some branched polymer theory, and we
will not discuss it here. Instead, we will take the double-trace
coupling constant to be some constant $\th$ in the large $N$
limit\foot{If we take the coupling constant to decrease as a power of
$N$, the diagrams involving the double-trace operators are negligible
in the large $N$ limit, so we get the same large $N$ string theory in
the zero string coupling limit.}.
Then, it is easy to see that the leading large $N$ limit will
be given by the sum of diagrams with $k+1$ spheres which are connected
only by $k$ double-trace vertex operators, giving a contribution which
scales as $\th^{k} N^2$ (see figure \diagrams). The subleading contributions
in $N$ will involve both non-planar diagrams and diagrams which are
connected to themselves by double-trace operators. The contribution of
any diagram may be simply computed by counting the number $k$ of
double-trace operators (corresponding to a $\th^k$ contribution) and then
replacing the double-trace operator by a small throat joining
smoothly the two worldsheets which it connects. A diagram which after this
replacement has genus $g$ will scale as $N^{2-2g}$ in the large $N$
limit\foot{It is easy to generalize this analysis also to vertices
involving a product of $n$ traces for higher $n$; the coefficients of
such vertices need to scale as $N^{2-n}$ in the large $N$ limit to
have a good (string-like) large $N$ expansion in which these vertices
affect the leading order terms in the genus expansion.}.

The full expression for the correlation functions in a theory which
has such a double-trace coupling will be of the form
\eqn\correltwo{\vev{\co_1 \co_2 \cdots \co_j} = \sum_{g=0}^{\infty}
N^{2-2g} \sum_{k=0}^{\infty} \th^k f_{g,k}(\lambda).}
Note that since we are adding a local double-trace operator, the two
worldsheets which are being connected by the vertex actually touch at
the position of the double-trace vertex in the four
dimensions along which the QFT lives\foot{We will limit ourselves here
to four dimensional field theories, though everything we say applies
to any dimension.}. Thus, it seems that
the new interactions
(governed by the perturbation theory in $\th$) should be realized as contact
interactions between worldsheets from the point of view of the string
theory describing the large $N$ limit of the field theory, as in
figure \twospheres.
As we will see below, in our AdS/CFT cases where the dual string theory lives
in more dimensions,
this picture will be modified.

Since the leading diagrams in the large $N$ limit are almost
disconnected, it is easy to compute them in terms of correlation
functions in the original theory before we added the double-trace
operator. For
example, suppose we add to a Lagrangian involving only single-trace
operators a term of the form
\eqn\double{\delta {\cal L} = \int d^4x {\th \over N^2}
\co_{n_1}(x) \co_{n_2}(x),}
for some single-trace operators $\co_{n_1}$ and $\co_{n_2}$.
Then, the large $N$ limit of any correlation
function will be of the form
\eqn\correlsplit{\eqalign{\vev{\co_1 \co_2 & \cdots \co_j} =
\vev{\co_1 \co_2 \cdots \co_j}_0 + \cr
& {\th \over N^2} \sum_{\rm partitions} \int d^4x \vev{\co_{i_1}
\co_{i_2} \cdots \co_{i_k} \co_{n_1}(x)}_0 \vev{\co_{n_2}(x)
\co_{i_{k+1}} \co_{i_{k+2}} \cdots \co_{i_j}}_0 + O(\th^2), \cr}}
where $\{ i_1,i_2,\cdots,i_k ; i_{k+1},\cdots,i_j \}$ is some
partition of the numbers from $1$ to $j$ into two groups, and
$\vev{}_0$ are the correlation functions in the theory before the
deformation. Note that all terms in \correlsplit\ scale as $N^2$ in
the large $N$ limit; there are also terms of higher order in $1/N^2$
which we did not write down explicitly.

\newsec{Explicit realizations of double-trace perturbations}

\noindent{\it (a) Running $\tilde h$}

Usually we do not discuss double-trace operators in Lagrangians since
(at least in four dimensions) they tend to be irrelevant, so we cannot
just add them to the Lagrangian. However, such terms can, and
generally do, arise when we
integrate out other fields.  For example, in the theories
studied in \refs{\tsezar,\adsil},
double-trace quartic scalar perturbations are generated radiatively in
QFTs arising at low energies on D-branes at codimension $<$ 6 orbifold
fixed planes. In this case the corresponding single-trace
perturbations are absent at large $N$ due to symmetries and
inheritance of ${\cal N}=4$ supersymmetric nonrenormalization
theorems. When double-trace operators arise from integrating out fields,
they necessarily arise with physical coefficients -- including
bare contributions plus possible counterterms -- scaling as
described in the previous section
(or smaller in the large $N$ limit, depending on
the genus of the diagram at which they first appear).
In these theories, there is a choice of QFT determined by the
specific renormalization condition chosen, which determines the
physical double-trace
coupling at a particular subtraction point $M$.  Different choices
are related by finite shifts in counterterms, which amount
to finite shifts in $\tilde h$ at scale $M$.

This case has the nice feature that there is a weakly-coupled
limit of the theory in which the double-trace interaction is
evident perturbatively \refs{\tsezar,\adsil}.
However, it has the complications of
broken supersymmetry and running couplings, and it is not clear
precisely what is the string theory dual of this case and how its
parameters are related to the field theory parameters. Therefore,
we will largely
focus on a strongly coupled but supersymmetric system in which
there is an exactly marginal double-trace perturbation.

\noindent{\it (b) The exactly marginal case}

We will
focus here on the following simpler case, where one can show that there
exists an exactly marginal double-trace operator. An advantage of
this case, beyond the fact that the operator is exactly marginal, is
that the single-trace operators involved are both chiral, so there is
no singularity when we bring them together and no subtlety in defining
the double-trace operator (which is just $\lim_{x\to y} \co_{n_1}(x)
\co_{n_2}(y)$). The example we will discuss is a deformation of the
one discussed by Klebanov and Witten in \klewit. Let us start from an
$SU(N)\times SU(N)$ $\cn=1$ supersymmetric gauge theory, with two
bifundamental chiral superfields $A_i$ ($i=1,2$) and two
anti-bifundamental chiral superfields
$B_j$ ($j=1,2$). This theory is believed to flow in the IR to a
strongly coupled fixed point; in fact, using the NSVZ formula for the
beta functions, one can show that for any value of the couplings the
beta functions for the two gauge couplings $g_1$ and $g_2$
are the same, so one
expects to have a one-dimensional surface of fixed points, which are
solutions to the equation $\beta(g_1,g_2)=0$. If we now deform the
theory by adding a superpotential
\eqn\suppot{W = N h\tr(A_1 B_1 A_2 B_2 - A_1 B_2
A_2 B_1)}
(which preserves the $SU(2)\times SU(2)\times U(1)_R$ global
symmetry of the original superconformal field theory), then one can show
\klewit\ by the methods of \ls\ that the beta function for $h$ is
proportional to that of $g_1$ and $g_2$, so that there is still just
one equation $\beta(g_1, g_2, h) = 0$ which needs to be satisfied for
the theory to be conformally invariant. Thus, we expect to have a two
dimensional manifold of exactly marginal deformations for these
conformal theories, and this was verified also in the AdS dual of
these theories in \klewit.

Now, we can ask what happens if we deform the theory described above
by an additional superpotential of the form
\eqn\nsuppot{\delta W = \th (\tr(A_1 B_1) \tr(A_2 B_2) - \tr(A_1 B_2)
\tr(A_2 B_1)),}
which also preserves the same global symmetries. It is easy to see
that the anomalous dimension of this operator is the same as that of
the operator in \suppot, so the beta functions of $h$ and $\tilde h$
are proportional to each other (as well as to the beta functions of
the gauge couplings). Therefore, we expect to still have a solution to
the equation $\beta=0$ even when we turn on the coupling $\th$, so
this is an exactly marginal coupling. In the scaling we wrote here the
large $N$ limit behaves as described in the previous section.
For small $\th$ we will
thus have in the large $N$ limit a double perturbation expansion in
$1/N^2$ and in $\th$, as described above\foot{We
could also have a perturbation expansion in
$h$, but in view of the AdS/CFT application we have in mind we prefer
not to take $h$ to be small, but just to be constant in the large $N$
limit. The supergravity approximation in this case is only valid at
large $h$.}.

Note that if we write the Lagrangian corresponding to \nsuppot\ in
components, the linear term in $\tilde h$ is indeed a sum of various
double-trace operators (including terms where each trace is either
bosonic of fermionic), but there is also the scalar potential (after
we integrate out the auxiliary fields) which includes a
triple-trace operator, whose coefficient is proportional to $|\th|^2 / N$
(where the $1/N$ comes from the K\" ahler metric in our
normalization). Such a term behaves in the large $N$ limit similarly to
the double-trace term we described in detail above (as would a term
with a product of $k$ traces and a coefficient $\th^{k-1}/N^{k-2}$).
We cannot write
down this term explicitly because we do not know the form of the
K\" ahler metric in this strongly coupled theory, but equation
\nsuppot\ uniquely specifies the deformation we are performing.

If we wish, we can alternatively describe this deformation by using an
auxiliary singlet field $\lambda_{ij}$ in the $\bf(2,2)$
representation of the global $SU(2)\times SU(2)$ symmetry group. Then,
we can write instead
\eqn\nnsuppot{\delta W = \epsilon^{ik} \epsilon^{jl}
(\lambda_{ij} \tr(A_k B_l) - {1\over {2\th}} \lambda_{ij} \lambda_{kl}).}
Integrating out the auxiliary field $\lambda_{ij}$ reproduces the action
\nsuppot. One can think of the contact interaction between
worldsheets described in the previous section
as arising from integrating out the infinitely massive auxiliary field
$\lambda_{ij}$.

For small $\th$ the correlation functions of the theory will
change by a small amount, given by \correlsplit\ (including
contributions from all the double-trace operators in our Lagrangian,
and at higher orders also from the triple-trace operator).
However, the moduli space of the theory changes in a more drastic way,
since some of it is lifted by the new term \nsuppot. The moduli space
of the original theory is \klewit\ a symmetric product of $N$ copies of the
conifold $z_{11} z_{22} - z_{12} z_{21} = 0$. On the moduli space we
can diagonalize all the matrices $A_1, A_2, B_1$ and $B_2$ (in some
arbitrary basis), and the gauge-invariant combinations (up to
permutations) are
$z_{ij}^{(I)} \equiv (A_i)_{II} (B_j)_{II}$ which can be identified
with positions on the conifold.

After we deform the theory, pairs of eigenvalues interact strongly
at long distance.  If only the
first eigenvalue of the matrices is non-zero, the potential still
vanishes. However, if two eigenvalues are non-zero, ${\del W\over \del
A_1}$ includes terms proportional to $\th (A_2)_{II} ((B_1)_{22}
(B_2)_{11} - (B_1)_{11} (B_2)_{22})$, which lead to
pairwise forces which grow with distance on
the conifold. One way to cancel all these forces is to impose
the four complex conditions $tr (A_iB_j)\equiv 0$. This suffices to
cancel the corrected F-terms, giving a branch of the moduli space
of real dimension $6N-8$ \foot{We thank D. Berenstein for
pointing out this branch of the moduli space.}. There is another
branch of the moduli space, of dimension $6+2(N-1)$, in which
this constraint does not necessarily hold, but
the $k$'th eigenvalue of all the matrices is given by $\alpha_k$
times the first eigenvalue for some complex number $\alpha_k$.

%The scalar potential
%including all contributions of this type
%vanishes if and only
%if $z_{ij}^{(2)} = \alpha z_{ij}^{(1)}$ for some (complex) constant
%$\alpha$. Similarly, we can only turn on additional eigenvalues if
%they are also proportional to the original ones. The moduli space is
%thus only $6+2(N-1)$ real-dimensional rather than being $6N$
%dimensional.

\newsec{Double-trace operators in the AdS/CFT correspondence}

\subsec{General properties}

The theories discussed in the previous section, before including the
double-trace deformation, have dual descriptions, using the AdS/CFT
correspondence \refs{\juan,\gkp,\witten,\review}, as type IIB string
theory on $AdS_5\times K$ for Einstein spaces $K$, where
$K=S^5/Z_n$ in the examples of \S3a,  and
$K=T^{1,1}$ in the example of \S3b \klewit\ (where $T^{1,1} =
(SU(2)\times SU(2) / U(1))$ is the base of the cone in
the conifold geometry). Thus, we can ask what these deformations correspond to
in the dual theory. Recall that in general, every single-trace
operator in the field theory corresponds to a field propagating on
$AdS$ space, and to a vertex operator in the corresponding
worldsheet string
theory. Deforming the field theory by a single-trace operator is dual
to turning on the corresponding vertex operator in the string theory,
or (equivalently) to looking at configurations in which the corresponding
field on $AdS$ has a particular behavior near the boundary. For
example, deforming the field theory of \S3b by the two exactly marginal
single-trace operators corresponds in the string theory dual
to changing the (constant over all
space) value of the string coupling and axion and of the integral of the
2-form fields over the non-trivial 2-cycle in $T^{1,1}$.

On the other hand, double-trace operators do not correspond to fields
on $AdS$ or to vertex operators in the string theory. In some sense
they correspond to two-particle configurations on $AdS$, since we can
define the double-trace operator by its appearance in the OPE of two
single-trace operators. This suggests that perhaps deforming by such
an operator should be described by some two-particle condensate in the
bulk (a ``squeezed state'') behaving in a particular way near the
boundary, analogous to the description of single-trace deformations.
One can perhaps construct such a state as a coherent state in an
off-shell formulation of the low-energy supergravity, but it is not
clear how to promote this description to the full string theory, which
does not contain any off-shell information in the bulk.
Thus, we will propose an alternative description for this deformation
based on (a generalization of) worldsheet string theory.

An important property of the double-trace deformation we are
discussing is that, unlike the single-trace deformations, the double-trace
deformation leads to a theory which is non-local in the ten dimensional
bulk :

\noindent
1. One way
to see the non-locality
is from the description above as a coherent state of two
particles, each of which is in some spherical harmonic on $K$;
obviously this induces long-term correlations between different
positions on $K$.

\noindent
2. The non-locality on $K$ also follows from the form of the
corrections \correlsplit\ to correlation functions. Correlation
functions of particular spherical harmonics will change in a different
way from those of other spherical harmonics; for example, only the two
point functions of the low-lying spherical harmonics corresponding to
the operators explicitly appearing in the double-trace term will change. This
again suggests that the deformation has no local description in ten
dimensions.

\noindent
3. Some of the original moduli space before we
turned on the deformation corresponded to configurations of D3-branes
sitting at arbitrary radial positions and arbitrary positions in
$K$. After the deformation, as described above, we can still
have a single D3-brane and no force will act on it, but if we have two
D3-branes at different positions on $K$ there should be a force
between them that grows with the distance (along $K$), coming from the
new scalar potential after the deformation. Again, this is
inconsistent with a local description in ten dimensions.
Formulas for similar long-range potentials can be found in \adsil\ for
case $(a)$.

\noindent
Note
that the scale of non-locality suggested by these observations is the
radius $R$ of $K$, which is also the radius of curvature of the
$AdS$ space; we will mostly be working in the limit where this is much
larger than the string scale (at which the theory is obviously non-local).
This requires us to consider case $(b)$ and
take the coupling $h$ discussed above to be large.

The arguments above suggest that after the deformation the theory is
non-local on $K$, but it could still be local on the $AdS$
space. The fact that the deformation is local in $\IR^4$ suggests that
it might be local also in $AdS$, but this is also consistent with a
non-locality of the order of the $AdS$ radius (since it is hard to
describe smaller objects than this in the field theory). In our
description below we will see that the resulting theory seems to be
non-local also on $AdS$.

\subsec{The deformation in string theory}

How can we describe the deformed theory ? Usually the only
deformations we are allowed to do in string theory have a
perturbative description involving turning on
vertex operators on the worldsheet.  In conformal perturbation
theory, one adds
\eqn\acshift{
\delta S = \epsilon \int d^2 \sigma \sqrt{\hat g} V
}
to the worldsheet action, which preserves conformal
invariance if $V$ is an exactly marginal physical vertex operator.
Working in the original theory and adding contributions
obtained by bringing down powers of
$\delta S$ into correlation functions produces the corrected
correlation functions of the deformed theory, perturbatively in $\epsilon$.
For example, if we consider a circle parameterized
by target space coordinate $X$, adding
$\int d^2\sigma\sqrt{\hat g} \epsilon V \equiv
\int d^2\sigma\sqrt{\hat g} \epsilon
\partial X\bar\partial X$
changes the radius squared of the circle by $\epsilon$.

More generally, one
can go beyond this conformal perturbation theory by re-solving
for the spectrum and interactions
of the deformed theory using the new
worldsheet action $S+\delta S$ (or equivalently by performing
the Polyakov path integral with the shifted action).
More generally still one can consider
terms of the form $\int d^2\sigma\sqrt{\hat g}
G_{\mu\nu}(X)\partial X^\mu \bar\partial X^\nu$
for which the condition for conformal invariance translates into
the condition $R_{\mu\nu}=0$ (at lowest order in $\alpha'$).

In the context of the AdS/CFT correspondence, these standard deformations
on the gravity side correspond only to deformations by
single-trace operators on the boundary, so the double-trace
deformation we are interested in here
cannot be described in this way. It seems that it is not
possible to describe this deformation from the point of view of a
single worldsheet, as suggested also by the field theory analysis
above where the deformation corresponds to a contact interaction
between worldsheets. We do not know how to describe the deformation at
a fundamental level, but we can reproduce the perturbation expansion
in $\th$ \correlsplit, analogous to the
perturbation theory about a fixed background worldsheet CFT deformed
by \acshift.
This is accomplished by adding an interaction term to the usual sum over all
worldsheets (connected and disconnected), as follows.

Suppose that our interaction in the gauge theory
is of the form \double, with $\co_{n_1}(x_1)$ of dimension $\Delta_1$
corresponding to a vertex operator of the form $V_{n_1}(\theta(w))
f_{\Delta_1,x_1}(x(w),z(w))$, where $V_{n_1}$ includes the appropriate
spherical harmonic as a function of the compact space, $w$
is the complex coordinate on the worldsheet\foot{When we write
$x(w)$ we refer to a general function/operator depending on $w$ and
$\bar w$.}, $x$ and $z$ are
coordinates on $AdS$ with the metric $ds_{AdS}^2 = {R^2\over z^2}(dz^2 +
dx^{\mu} dx_{\mu})$, and
$f_{\Delta_1,x_1}$ is the non-normalizable wave-function on $AdS$ for
an operator of dimension $\Delta_1$ with delta-function support at
a point $x_1$ on the boundary.
There is a similar vertex operator corresponding to $\co_{n_2}(x_2)$.
We are being schematic here since in
any case we do not have a good description of string theory in this RR
background.
Then, it seems that we
need to add an interaction between two worldsheets of the form
\eqn\twosheet{\eqalign{\delta {\tilde S} =
\th \int d^4x &\int d^2w_1 V_{n_1}(\theta_1(w_1))
f_{\Delta_1,x}(x_1(w_1), z_1(w_1)) \cdot \cr
\cdot &\int d^2w_2 V_{n_2}(\theta_2(w_2))
f_{\Delta_2,x}(x_2(w_2),z_2(w_2)) \equiv \cr
\equiv \th \int d^2&w_1 \int d^2w_2 V_{n_1}(\theta_1(w_1))
V_{n_2}(\theta_2(w_2))
K(x_1(w_1),z_1(w_1);x_2(w_2),z_2(w_2)), \cr}}
where
\eqn\kernel{K(x_1,z_1;x_2,z_2) = \int d^4x G_{\Delta_1}(x; x_1,z_1)
G_{\Delta_2}(x; x_2,z_2),}
and $G_{\Delta}$ is the boundary-to-bulk propagator on $AdS_5$, given in
Euclidean space by
\eqn\prop{G_{\Delta}(x; x_1,z_1) = \pi^{-2} {{\Gamma(\Delta)}\over
 \Gamma(\Delta-2)} {{z_1}^\Delta \over (z_1^2 + (x-x_1)^2)^\Delta}.}
As in our field theory discussion, $w_1$ and $w_2$ can either be on
the same connected component of the worldsheet or on different
connected components.
For two points on the boundary, $z_1=z_2=0$, $K$ is just a delta
function (as expected from the field theory analysis of \S2,
which suggested a contact interaction in 4d),
but in the bulk it is non-zero also when $x_1 \neq x_2$.
Our vertex operators appearing in \twosheet\
contain an implicit factor of the string coupling $g_s$, as is
standard for vertex operators describing ordinary string excitations
(though usually one does not include this factor in the
standard deformations \acshift\ describing condensation of
strings).
This yields the correct $g_s$-dependence to match the field
theory expansion \correltwo\double\correlsplit\ (with the normalization
of field theory operators as discussed in the previous sections).
In momentum space we can write $K$ as
\eqn\kpropE{K(x_1,z_1;x_2,z_2) = \int d^4k e^{ik\cdot (x_1-x_2)} z_1^2
z_2^2 K_{\Delta_1-2}(|k|z_1) K_{\Delta_2-2}(|k|z_2)}
in Euclidean space; the Lorentzian case is defined as usual by
analytic continuation of this expression, which involves the same
modified Bessel functions $K_{\nu}(z)$ for $k^2 > 0$ and
Hankel functions $H_{\nu}(z)$ for $k^2 < 0$
\BalasubramanianSN.
%, or in the form
%\eqn\kpropL{\eqalign{K(x_1,z_1;x_2,z_2) = &
%\int_{k^2 > 0} d^4k e^{ik\cdot (x_1-x_2)} z_1^2
%z_2^2 K_{\Delta_1-2}(|k|z_1) K_{\Delta_2-2}(|k|z_2) + \cr
%&\int_{k^2 < 0} d^4k e^{ik\cdot (x_1-x_2)} z_1^2
%z_2^2 A_{ab} H_{\Delta_1-2}^{(a)}(|k|z_1) H_{\Delta_2-2}^{(b)}(|k|z_2) \cr}}
%in the Lorentzian case \BalasubramanianSN, where $K_{\nu}(z)$ and $H_{\nu}^
%{(1)}(z)$ are
%modified Bessel functions and Hankel functions, respectively, and
%$A_{ab}$ is hermitian.
%As usual, the Lorentzian theory is defined by analytic continuation of
%the Euclidean case.

The added interaction \twosheet\ manifestly reproduces the leading term in
\correlsplit, at least in the supergravity approximation.
Since the interaction \twosheet\ that we added is a sum of products of
vertex operators (of dimension $(1,1)$)
on each worldsheet, it is clear that it preserves
conformal invariance.
However, since this interaction
involves two worldsheets, it does not seem to be
equivalent to any standard string interaction which is local on the
worldsheet.

Similarly, we can
generalize \twosheet\ to an interaction between three strings that
would correspond to a triple-trace operator, which we need to do if we
want to add a double-trace operator in a way which preserves
supersymmetry (as described in the previous section).
Note that (as argued, for instance, in \gk) when we compute sphere
correlation functions on $AdS$,
the volume of $SL(2, C)$ is absorbed by the integral
over the radial position on $AdS$, so two-point functions do not
necessarily vanish on the sphere.

There is one simplification evident from
the diagrammatics of our deformation that is worth
pointing out.
Although this deformation is not a modulus
of local supergravity, and has various novel
nonlocal features as discussed in \S4.1,
it follows from our perturbative formulation in this section
that graviton scattering alone is unaffected by
the deformation at order ${\cal O}(N^2)$ (tree-level
on the gravity side).  At this order, all diagrams
with only gravitons on external legs
involve at least one genus zero component
of the worldsheet with insertions of a single
field $\phi_{n_1}$ or $\phi_{n_2}$ (dual
to the factors ${\cal O}_{n_{1,2}}$ in the
double-trace perturbation), combined with some
number of gravitons.  At tree level, matter
fields do not couple linearly to gravitons, and
these diagrams all vanish.  In the case of the
orbifold models of case (a), in fact
the self-interactions of all untwisted
modes are unaffected by the deformation at ${\cal O}(N^2)$.
This follows from the inheritance of untwisted amplitudes
in orbifold field and string theories at this order.

\subsec{Comments on possible generalizations}

Next, we can ask if, as in the case of
ordinary single vertex operator
deformations whose condensation leads to a change in the space-time
background appearing in the worldsheet Lagrangian,
we can find
also in this case an exact
description going beyond our generalization of worldsheet conformal
perturbation theory. Perturbatively, we describe our deformation by a
Polyakov path integral of the form
\eqn\genpol{\sum_{disconnected}\int  [DY] e^{-S_0-\delta {\tilde S}},
}
where $S_0$ is the worldsheet action before the deformation,
and where
we use $Y\equiv (\theta, x, z)$ to denote the full set of
space-time coordinates.
We would like to
know if
we can describe the deformation more generally using
a Polyakov path integral of the form
\eqn\ngenpol{\sum_{disconnected}\int  [DY] e^{-S},}
where
\eqn\genac{\eqalign{
S=&\sum_{I=1}^{N_w} \int d^2 \sigma^{(I)} \sqrt{\hat g(\sigma)}{\cal L}_0
~~ + ~~ \cr
&\sum_{I=1}^{N_w} \int d^2\sigma_1^{(I)}\sqrt{\hat g(\sigma_1)}
\sum_{J=1}^{N_w} \int d^2\sigma_2^{(J)}\sqrt{\hat g(\sigma_2)}
{\hat K}[Y(\sigma_1), Y(\sigma_2)]
\cr
&+~~trilocal ~ and ~ higher ~ contributions}}
for some function $\hat K$,
where we wrote the worldsheet coordinate $\sigma$ as
a direct sum of contributions
from different connected components of the
worldsheet:  $\sigma = \sum_{I,\oplus} \sigma^{(I)}= \sigma^{(1)}
\oplus \sigma^{(2)} \oplus \dots\oplus \sigma^{(N_w)}$,
and $N_w$ is the number of disconnected components of the
worldsheet in a given term of \ngenpol\foot{Similar non-local terms in the
worldsheet action
were studied on connected worldsheets in \willy\ in analyzing
the problem of D-brane recoil.}.  We have included in \genac\
the possibility of trilocal and higher multilocal terms on
the worldsheet, which may be required to cancel violations of
Weyl invariance that arise as operators from different
bilocal terms in \genac\ collide on a given
component of the worldsheet\foot{Alternatively, we can think of such terms
as arising in a worldsheet
renormalization group flow starting from a theory with
only bilocal couplings.}.
If this description is to work, we need to know
what the conditions on the couplings in $S$ are in order to
have conformal invariance.  These should follow from requiring
cancellation of the anomaly under Weyl rescalings
of the worldsheet metric (${\hat g}_{\alpha\beta}\to
e^{2\eta}{\hat g}_{\alpha\beta}$).
In other words, if we
calculate the Weyl anomaly for the metric using the action \genac,
what conditions do we get on
${\hat K}[Y(\sigma_1), Y(\sigma_2)]$ and on the higher nonlocal terms
in \genac ?
In the linearized approximation one solution for ${\hat K}$
is \twosheet-\kpropE,
which appears to reproduce
the double-trace deformation in conformal perturbation theory.
More generally, we might expect to find a much wider class of solutions,
analogous to the solutions of the Ricci-flatness condition
in the case of the Lagrangian $G_{\mu\nu}(X)\partial X^\mu
\bar\partial X^\nu$
on a single worldsheet.  Most of these solutions
will not involve products of propagators for single quanta of the
bulk fields as we had above,
but will instead involve much more general
field configurations ${\hat K}(Y_1,Y_2)$ solving the conformal
invariance conditions. Note that in this formalism we still have the
usual genus expansion (including also disconnected worldsheets), but
with a non-local worldsheet action which relates (possibly
disconnected) components of the worldsheet, and changes the power of
$g_s$ associated with certain diagrams.

On D-brane probes, there will be new interactions obtained from
worldsheets of the sort discussed above but with boundaries on
the D-brane probes.  In the worldvolume theory, these will
in general appear as new interactions.  It would be interesting
to obtain the rules for what sorts of couplings can consistently
be introduced on D-brane probes; perhaps this can be used
to constrain the
general possibilities for such interactions on the string theory side.

\newsec{An alternative description of the deformation}

An equivalent way to describe the deformed theory is by using our
description \nnsuppot\ of the deformation in the field theory. If we
had only the first term in \nnsuppot, then the auxiliary field
$\lambda_{ij}$ appearing there would be identified in string theory with
the boundary value for a non-normalizable mode of the field
corresponding to the operator $\tr(AB)$. Thus, we can describe the
deformed theory by integrating over all possible boundary conditions
for this field, with a weight given by $e^{-{1\over 2\th} \int
\lambda_{ij}^2}$ (the precise description is actually
a bit more complicated because
\nnsuppot\ is a superpotential and not a Lagrangian, but this raises
no new issues). From the bulk point of view, this description
may be less useful than the description above, since it involves
integrating over boundary conditions and completely obscures the
physics in the bulk.
However, if we do a perturbation expansion of this
description, we recover the description above, where $K$ essentially
arises as the inverse propagator for $\lambda_{ij}$ in the bulk.

If we discuss a theory on an $AdS$ space with a cutoff (which is a
UV cutoff in the field theory), then all the couplings, including
$\lambda_{ij}$, become dynamical fields. We can then introduce interactions
like the second term of
\nnsuppot\ on the cutoff, which will lead to the theories we
described above as we take the cutoff to infinity (the kinetic term of
$\lambda_{ij}$ goes to zero in this limit).

One way to view the new interactions is, therefore, in terms of adding
an auxiliary field $\lambda_{ij}$ on the boundary,
such that the new interactions all
involve this field. Thus, it might seem that we are really not
changing the bulk physics at all. However, from the discussion of the
previous sections it is clear that (for instance) from the point of
view of the low-energy theory the couplings in the bulk change, just
like for single-trace deformations.  In particular, the double-trace
perturbation we are discussing in the conifold case is an exactly
marginal perturbation on the field theory side, which therefore
affects physics on all scales, and in the cases of \S3a\ the
dynamically generated double-trace contribution grows in the
infrared of the field theory. Therefore, on the gravity side
these deformations should have effects in the bulk of the
AdS space rather than being concentrated at the boundary.
Also, one might think that two
worldsheets will only interact if a particle can actually be physically
transmitted from one to the other through the boundary, but this is
clearly not true;  there is an infinite potential barrier preventing
most normalizable particle excitations in the interior of the space from
propagating to the boundary.  The bulk-boundary propagator
involves instead nonnormalizable excitations whose wavefunctions near
the boundary are infinitely rescaled relative to those of normalizable
excitations, as discussed for example in
\refs{\GiddingsQU,\PolchinskiRY} in the
context of the undeformed theory.
Furthermore, from the form of \kernel\ it seems clear that the new
interactions relate also nearby points in the bulk, which cannot be
causally connected through the boundary\foot{This arises at least
in part from the fact
that the propagators appearing in the Lorentzian continuation of
\kpropE\ have exponentially decaying
contributions outside the light cone.  This is of course also a feature
of Feynman propagators in ordinary quantum field theory.  Here, as
there, to diagnose whether acausal effects really arise one would need to
study more refined observables such as expectation values of commutators
of spacelike separated fields in the bulk.}.

It would be interesting to understand if the choice of boundary
conditions and path integral weights that we are making
can be related to a choice made
in taking the near-horizon limit of the full D-brane systems leading
to the AdS/CFT dual pairs we are studying.  In taking the near-horizon
limit, the non-normalizable modes which live in the asymptotically
flat region away from the brane freeze out, serving as sources
(couplings) in the field theory and background parameters in the
gravity theory.  In the standard limit \juan, the massive fields among
them are simply set to zero.  But perhaps the near-horizon limit is
more subtle in general; it is tempting to speculate that the $\lambda_{ij}$
appearing in the above prescription are some of these physical
asymptotic closed string states, and that there is perhaps some other
way of taking the near-horizon limit in which these states,
while still not corresponding to dynamical fields,
affect the theory differently from what one
would get by simply setting them to zero, because of the residual
couplings \nnsuppot.

\vskip 1cm

\centerline{\bf Acknowledgments}

We would like to thank A. Adams, T. Banks, D. Berenstein, S. Elitzur,
S. Giddings, D. Gross, A. Hashimoto, S. Kachru,
I. Klebanov, D. Kutasov, J. Maldacena, H. Ooguri, Y. Oz, M. R. Plesser,
J. Polchinski, E. Rabinovici, N. Seiberg,
S. Shenker, and H. Verlinde for interesting and useful discussions.
Our work was supported in part by a grant from the
United States-Israel Binational Science
Foundation (BSF), and by the Institute of Theoretical Physics
at UCSB where this project was initiated.  The work of O.A. and
M.B. was also supported by the IRF Centers of Excellence program, by
the European RTN network HPRN-CT-2000-00122, and by Minerva.
The work of E.S. was also supported by the DOE under contract
DE-AC03-76SF00515 and via an OJI grant, and by a Sloan fellowship.

\listrefs

\end